\documentclass{iau-JDSS}
\usepackage{graphicx}

\title[Diffuse interstellar bands]{New views on the diffuse interstellar bands}

\author[N.L.J. Cox]{Nick L.J. Cox$^1$}

\affiliation{$^1$Institute of Astronomy, KU Leuven, Leuven, Belgium \break
(presently at IRAP/CNRS, Toulouse, France - nick.cox@irap.omp.eu)}

\pubyear{2015}
\volume{Volume 16}
\pagerange{}
\date{}
\jname{Highlights of Astronomy, Volume 16}
\editors{Thierry Montmerle}
\begin{document}

\maketitle

\begin{abstract}
New views on the diffuse interstellar bands are discussed. In particular results from DIB surveys 
and the study of near-infrared DIBs.
\end{abstract}

Heger (1922) was the first to establish the interstellar origin of the diffuse interstellar bands (DIBs). Later, 
Merrill \& Wilson (1938), showed that DIBs correlate with both interstellar gas and dust. Recent surveys confirm that DIBs correlate with both gas and dust (e.g. Friedman et al. 2011, Vos et al. 2011). The latest tally of DIBs stands at over 400 (Hobbs et al. 2009) of which the majority are weak ($<$~20~m\AA/E$_{(B-V)}$) and narrow ($\sim$~1~\AA). There are only a dozen or so strong ($>$~100~m\AA/E$_{(B-V)}$) narrow/broad DIBs. It is clear that DIB carriers constitute a significant component of the diffuse ISM. The total equivalent width of the DIBs suggests a total carrier abundance of 10$^{14}$~cm$^{-2}$ (Cox 2011). In particular the change in relative strengths between DIBs in different sightlines suggests that there is a direct impact of the environmental conditions, providing potential for diagnostic use. Also, none but two of the strong DIBs are strongly correlated with each other (McCall et al. 2010). 
Cami \& Cox (2014) provide an overview of recent advances in observational, theoretical and laboratory work on DIB carriers. A road map for the identification of the DIB carriers is presented in Cox \& Cami (2014). Here I will focus on new results from DIB surveys and searches for near-infrared bands. 

\section{New views on the diffuse interstellar bands}\label{sec:new}

\subsubsection*{Galactic and extra-galactic surveys}
Vos et al. (2011) made a survey of 78 lines-of-sight in the Upper Scorpius region. This work revealed a relation
between the 5797/5780 DIB strength ratio and the local effective interstellar radiation field. This corroborates
the so-called $\sigma$-$\zeta$ effect, which postulates that the relative strength of some DIBs is due to
ionisation cq. destruction of DIB carriers at the edges of diffuse clouds with respect to those in the interior of
clouds. Surveys of Galactic lines-of-sight with HERMES at the Mercator telescope are also on-going in anticipation
of the observation of the 8621~\AA\ DIB with the Gaia astrometry mission. In particular we focus on its intrinsic 
profile and its relation to visual DIBs. Similar studies of the ISM in nearby galaxies are now becoming feasible.
The DIB strengths in M31 are related in similar fashion to reddening as in Galactic environments. No significant
correlation was found with strength of the interstellar radiation field nor with PAH emission (Cordiner et al. 2011). 
The formation and destruction of DIB carriers depends primarily on the local physical conditions, which prevail in 
most, if not all, galaxies.

\subsubsection*{Near-infrared DIBs}
PAHs are believed to be abundant in space, as inferred from the ubiquitous presence
of the mid-infrared emission bands, the extended red emission and the UV bump (see also contribution from Joblin,
Vijh, Mulas, Jones, Li, and Kwok in this volume). The presence of PAHs in the ISM leads naturally to the
hypothesis that their electronic absorption spectra could be detected. This is the basis of the PAH-DIB hypothesis
(e.g. Cox 2010). Small ionised species absorb in the visual, with the main electronic transitions progressively shifting towards longer (near-infrared) wavelengths with increasing size. 

Previously, few near-infrared DIBs were known. Joblin et al. (1990) reported two DIBs at 11797 and 13175~\AA.
Foing \& Ehrenfreund (1994) detected two additional near-infrared DIBs at 9577 and 9632~\AA, which they attributed tentatively to C$_{60}^+$. Only recently, an additional 13 near-infrared DIBs (between 1.5 to 1.8~$\mu$m) have been identified towards the Galactic Centre (Geballe et al. 2011). New X-Shooter spectra have been obtained for sightlines probing visual extinctions from $\sim$~1 to 10~mag. These observations confirm all but the faintest DIBs reported by Geballe, and show that indeed their strengths correlate roughly with visual extinction. First results show that the 15268~\AA\ DIB correlates well with the 5780~\AA\ DIB, but not with the 5797~\AA\ DIB (Cox et al. 2014).

\section{Future work on DIB surveys and the near-infrared DIBs}

DIB surveys are necessary to address questions regarding the depletion/absence of DIBs, relative to the gas and dust content, in certain astrophysical environments, and to study environments in which DIB properties and behaviour are altered. Results from such surveys can then put new constraints on their carriers. Furthermore, DIBs constitute a (yet unknown) component of the ISM, which can already be used as an environmental diagnostic. This is particularly useful for studying extra-galactic diffuse medium.

A complete inventory of near-infrared DIBs is still lacking. The properties (profile sub-structure, asymmetry, etc) and
behaviour of the near-infrared DIBs needs to be studied in detail to understand their link with the visual DIBs as well as their link with PAHs and related molecules. It would be interesting to see if and how near-infrared DIBs connect with other 
mid-infrared ``organic'' spectral features, such as the 3.3~$\mu$m band.

\begin{acknowledgments}
I thank FWO for financial support to attend this meeting. I also acknowledge support from the Belgian Federal Science Policy Office via the PRODEX Programme of ESA.
\end{acknowledgments}

\end{document}